\documentclass[prd,aps,amssymb,nofootinbib,showpacs,preprint]{revtex4}
\usepackage{amsmath,amsfonts}
\usepackage{hyperref}
\usepackage{graphicx}
\usepackage{epstopdf}

\begin{document}

\title{Scalar self-force on static charge in a long throat}

\author{A. Popov}
\address{Kazan Federal University, 18 Kremlyovskaya St., Kazan 420008, Russia\\
apopov@ksu.ru}

\author{O. Aslan}
\address{Kazan Federal University, 18 Kremlyovskaya St., Kazan 420008, Russia\\
alsucuk@gmail.com}

\begin{abstract}
We compute the self-force on a scalar charge at rest in the spacetime of long throat. We consider arbitrary values of the mass of the scalar field and the constant of nonmnimal coupling of the scalar field to the curvature of spacetime.
We also show the coincidence of explicit calculations of self-force in the limit of large mass of the field with known results.

\end{abstract}


\maketitle

\section{Introdution}

A very well known phenomenon that occur with a charge in a curved spacetime, is that it may become subjected to the self-interactions. The origin of this induced self-interaction resides on the non-local structure of the field caused by the spacetime curvature or non-trivial topology.

In flat spacetime this effect is produced by a local distortion of the field lines associated with the particle's acceleration.
For electrically charged particles in flat spacetime, the self-force is given by the Abraham-Lorentz-Dirac formula\cite{Dirac:1938,Poisson:1999}.
In the gravitational field the self-energy problem becomes more complicated. The reason is that contribution to the self-energy in this case is non-local.
The self-force problem for an electric charge in a curved space background was first investigated by DeWitt and Brehme\cite{DeW-Bre:1960} and later by Hobbs\cite{Hobbs1:1968}.
The gravitational self-force was first calculated almost simultaneously by Mino, Sasaki and Tanaka\cite{MST:1997} and by Quinn and Wald\cite{Quinn:1997}.  Later, Quinn derived the equivalent formula for a charge coupled to a minimally-coupled massless scalar field\cite{Quinn:2000}.

A number of simple static configurations has been analyzed, including the self-force acting on scalar or electric charges held static in the spacetime of a Schwarzschild black hole\cite{SmithWill:1980,ZelFrol:1982,Wiseman:2000,Burko:cqg}, electric or magnetic dipoles which are static outside a Schwarzschild black hole\cite{LeauteLinet:1984}, a static electric charge outside a Kerr black hole\cite{LeauteLinet:1982,OttewillTaylor:2012} or a Kerr-Newman black hole\cite{Lohiya:1982}, a static electric charge in a spherically-symmetric Brans-Dicke field\cite{LinetTeyssandier:1979}.
The self-force can be nonzero for a static particle in flat spacetimes of the topological defects\cite{Linet:1986,Smith:1990,Khus:1994,Khus:1995,deLor:2002}. In curved spacetimes with nontrivial topological structure the investigations of this type have the additional interesting features\cite{KhusBakh:2007,Linet:2007,Kra:2008,BezKhu:2009,Casals:2009,Popov:2010}.

Unfortunately, the authors do not know the results of calculation of the self-force of the charge, which is the source of a massive field.
In this paper we consider the problem of computing the self-force on a scalar charge at rest in the spacetimes of long throats, allowing for the arbitrary values of the mass of field and coupling constant. It gives the possibility to compare the explicit calculation of the self-force in the limit of large mass of the field with the corresponding result of paper\cite{Popov:2011}.

Throughout this paper we use units $c=G=1$.

\section{WKB approximation of the self-potential}

Let us consider a scalar field $\phi$ with scalar source $j$. The corresponding
field equation has a form
\begin{equation} \label{meq}
{\phi}^{;\mu}_{;\mu} - (\xi R+m^{2})  \phi = -4\pi q \int
\delta^{(4)}(x^\mu,\tilde x^\mu(\tau)) \frac{d\tau}{ \sqrt{-g^{(4)}}},
\end{equation}
where $\xi$ is a coupling of the scalar field to the scalar curvature
$R$ and $g^{(4)}$ is the determinant of the metric $g_{\mu \nu}$, $m$ denotes the mass of the scalar field, $q$ is the scalar charge and $\tau$ is its proper time.
The world line of the charge is given by $\tilde x^\mu(\tau)$.
We shall consider the case in which the charge is at rest in a static spacetime
\begin{equation} \label{metric0}
ds^2= g_{tt}(x^i)d t^2+g_{j k}(x^i)d x^j d x^k,
\end{equation}
where $i, j, k = 1, 2, 3$.
This means that one can rewrite the field equation in the
following way
        \begin{eqnarray} \label{beq0}
        \frac{1}{\sqrt{-g_{tt}}\sqrt{g^{(3)}}} \frac{\partial}{\partial x^j}
        \left( \sqrt{-g_{tt}}\sqrt{g^{(3)}} g^{j k} \frac{\partial
        \phi(x^i; \tilde x^i)}{\partial x^k}   \right)
        \nonumber \\
        -(\xi R(x)+m^{2}) \phi(x^i; \tilde x^i)
        =-\frac{4 \pi q \delta^{(3)}(x^i, \tilde x^i)}{\sqrt{g^{(3)}}},
        \end{eqnarray}
where $g^{(3)}=\det g_{i j}$ and we take into account that $d\tau/dt=\sqrt{g_{tt}}$ for the charge at rest.

In the static spherically symmetric spacetime
\begin{equation}\label{metric}
ds^2= -f(\rho) d t^2+d\rho^2+r^2(\rho)\left(d\theta^2+\sin^2\theta\, d\varphi^2 \right)
\end{equation}
one can rewrite the equation (\ref{beq0}) in the following way
\begin{eqnarray}
&&\left[ \frac{\partial^2}{\partial \rho^2} +\left( \frac{{f}'}{2 f}+\frac{(r^2)'}{r^2} \right)
\frac{\partial}{\partial \rho}+\frac{\partial^2}{\partial \theta^2}+\cot \theta \frac{\partial}{\partial \theta}
\right. \nonumber \\ && \left.
+\frac{\partial^2}{\partial \varphi^2}
-(\xi R+m^{2}) \frac{}{}\right]
\phi(\rho, \theta, \varphi; \tilde \rho, \tilde \theta, \tilde \varphi)
\nonumber \\ &&
=-\frac{4 \pi q \delta(\rho, \tilde \rho) \delta(\theta, \tilde \theta) \delta(\varphi, \tilde \varphi)}{r^2 \sin \theta}.
\end{eqnarray}
The solution of this equation can be expanded in terms of Legendre polynomials $P_l$ with the result that
\begin{equation} \label{phi}
\phi(x^\alpha; \tilde x^\alpha)= q \sum_{l=0}^\infty
\left(2l+1\right) P_l(\cos\gamma) g_l(\rho,\tilde \rho),
\end{equation}
where
$\cos \gamma \equiv \cos \theta \cos \tilde \theta
+\sin \theta \sin \tilde \theta \cos(\varphi-\tilde \varphi)$
and $g_l(\rho,\tilde \rho)$ satisfies the equation
\begin{eqnarray} \label{gl}
&& g_l'' + \left( \frac{{f}'}{2 f}+\frac{(r^2)'}{r^2} \right)g_l' - \left[ \frac{l(l+1)}{r^2} +m^{2}+ \xi
R\right]g_l
= -\frac{\delta(\rho, \tilde \rho)}{r^2}.
\end{eqnarray}
In this expression and below a prime denotes a derivative with respect to $\rho$. The homogeneous solutions to this equation will be denoted by $p_l(\rho)$ and $q_l(\rho)$. $p_l(\rho)$ is chosen to be the solution which is well behaved at $\rho =-\infty$ and divergent at $\rho \rightarrow +\infty$. $q_l(\rho)$ is chosen to be the solution which is divergent at $\rho \rightarrow -\infty$ and well behaved at $\rho =+\infty$. Thus
\begin{eqnarray} \label{pqeq}
&& \left\{\frac{d}{d \rho^2} + \left( \frac{{f}'}{2 f}+\frac{(r^2)'}{r^2} \right)\frac{d}{d \rho} - \left[ \frac{l(l+1)}{r^2} +m^{2}
 + \xi R\right]\right\}\left\{p_{\, l}(\rho) \atop q_l(\rho) \right\} = 0,
\end{eqnarray}
\begin{eqnarray}
g_l(\rho,\tilde \rho)&=& C_l p_{\, l}(\rho_<) q_l(\rho_>)
\nonumber \\
&=& C_l \left[\frac{}{} \Theta (\tilde \rho-\rho) p_{\, l}(\rho) q_l(\tilde \rho)
+\, \Theta (\rho-\tilde \rho) p_{\, l}(\tilde \rho) q_l(\rho)\right],
\end{eqnarray}
where $\Theta(x)$ is the Heaviside step function, i.e.,
$\Theta(x) = 1$ for $x > 0$
and $\Theta(x) = 0$ for $x < 0$,
$C_l$ is a normalization constant which could
be absorbed into the definition of $p_{\, l}$ and $q_l$. Normalization of $g_l$ is
achieved by integrating (\ref{gl}) once with respect to $\rho$ from
$\tilde \rho - \delta$ äî $\tilde \rho + \delta$ and letting $\delta
\rightarrow 0$. This results in the Wronskian condition
\begin{equation} \label{WC}
C_l \left( p_{\, l} \frac{d q_l}{d \rho}- q_l \frac{d p_{\, l}}{d \rho} \right)
=-\frac{1}{r^2}.
\end{equation}

The WKB approximation for the radial modes $p_{\, l}$ and $q_{\, l}$ is obtained
by the change of variables
\begin{eqnarray} \label{pq}
p_{\, l}&=&\frac{1}{\sqrt{2r^2W}} \exp\left({\int^{\rho}} W d \rho \right),
\nonumber \\
q_l&=&\frac{1}{\sqrt{2r^2W}} \exp \left({-\int^{\rho}} W  d \rho \right).
\end{eqnarray}
Substitution of these expressions into (\ref{WC}) shows that the Wronskian
condition is obeyed if
\begin{equation}
C_l=1.
\end{equation}
Substituting the expressions (\ref{pq}) into (\ref{phi}) we can obtain the following expression for $\phi$ under the assumptions $\theta=\tilde\theta, \varphi=\tilde \varphi$ and $\tilde \rho=\rho+\delta \rho > \rho$
\begin{eqnarray} \label{phi2}
&&\phi(\rho, \theta, \varphi; \tilde \rho, \theta, \varphi)
=\frac{q}{r(\rho) r(\tilde \rho)}
 \sum_{l=0}^\infty \left(l+\frac12\right)
\frac{\exp \left({-\int\limits^{\rho+\delta \rho}_{\rho}}
{ W \left(\tilde{\tilde \rho}, l+\displaystyle \frac12\right)} d \tilde{\tilde \rho} \right)}
{ \sqrt { W \left(\rho, l+ \displaystyle \frac12\right) W \left(\tilde \rho, l
+\displaystyle \frac12\right)}}.
\end{eqnarray}
Substitution the expressions (\ref{pq}) into the mode equation (\ref{pqeq}) gives the following equation for $W$:
\begin{eqnarray} \label{Weq}
&&W^2={ \frac{l(l+1)+m^{2}r^{2}+2 \xi }{r^2}}+\frac{{\left( W^2 \right)}''}{4 W^2}
-\frac{5 {{\left( W^2 \right)}'}^2}{16 W^4}
+\frac{{ f'}{(W^2)}'}{8 f W^2}-\frac{ {f}' W }{2 f}
+\frac{{(r^2)}''}{2 r^2}
\nonumber \\ &&
-\frac{{{(r^2)}'}^2}{4 r^4}+\frac{{(r^2)}' f'}{4 r^2 f}
+\xi\left( -2\frac{{(r^2)}''}{r^2}+\frac{{{(r^2)}'}^2}{2 r^4}-\frac{{(r^2)}' f'}{r^2 f} -\frac{f''}{f}+\frac{{{f}'}^2}{2 f^2}\right).
\end{eqnarray}
This equation can be solved iteratively when the metric functions $f(\rho)$ and  $r^2(\rho)$ is slowly varying, that is,
     \begin{equation} \label{lwkb}
     \varepsilon_{\mbox{\tiny \sl WKB}}=L_{\star} /L \ll 1,
     \end{equation}
where
      \begin{equation} \label{Lst}
      L_{\star}(\rho) =\frac{r(\rho)}{\sqrt{2\xi+m^{2}r^{2}(\rho)}},
      \end{equation}
and $L$ is a characteristic scale of variation of $f(\rho)$ and  $r^2(\rho)$:
       \begin{eqnarray} \label{Lm}
       &&\frac{1}{L(\rho)}= \max \left \{ \left| \frac{r'}{r}  \right|, \
       \left| \frac{f'}{f}  \right|, \
       \left| \frac{r'}{r} \sqrt{\left|\xi \right|} \right|, \
       \left| \frac{f'}{f} \sqrt{\left|\xi \right|} \right|, \
       \left| \frac{r''}{r}  \right|^{1/2}, \
       \left| \frac{f''}{f}  \right|^{1/2}, \
       \dots  \right \} .
       \end{eqnarray}
We shall call the region of spacetime where the metric functions $f(\rho)$ and  $r^2(\rho)$ is slowly varying \textsl{the long throat}. This type region exists, for example, in the neighborhood of the ultraextreme  horizon\cite{PopZas:2007}.

The zeroth-order WKB solution of Eq. (\ref{Weq}) corresponds to neglecting
terms with derivatives in this equation
       \begin{equation} \label{Wsol}
       W^2=\Omega 
       \cdot
       \left(\frac{}{}1+O(\varepsilon_{\mbox{\tiny \sl WKB}})  \right),
       \end{equation}
where
        \begin{eqnarray}
        &&\Omega\left(\rho, l+1/2\right)=\frac{l(l+1)+m^{2}r^{2}+2 \xi }{r^2}
        =\frac{1}{r(\rho)^2}\left[
        \left(l+\frac12\right)^2+\mu^2\right],
        \end{eqnarray}
and
        \begin{equation}
       \mu^2=2\xi-\frac14+m^{2}r^{2}.
        \end{equation}
Below it is assumed that
       \begin{eqnarray}
       \mu^2 > 0.
       \end{eqnarray}

Let us stress that $\Omega$ is the exact solution of Eq. (\ref{Weq} in a spacetime
with metric
$ds^2=-f_0 d t^2+d\rho^2+r_0^2(d\theta^2+\sin^2\theta\, d\varphi^2)$,
where $f_0, r_0$ are constants.

Substituting the solution (\ref{Wsol}) into (\ref{phi2}), and neglecting terms of the first order and higher with respect to $\varepsilon_{\mbox{\tiny \sl WKB}}$ we can obtain
\begin{eqnarray}
&&\phi(\rho, \theta, \varphi; \tilde \rho, \theta, \varphi)=
\frac{q}{r(\rho) r(\tilde \rho)}
\sum_{l=0}^\infty \left(l+\frac12\right)
\frac{\exp \left({-\int\limits^{\rho+\delta \rho}_{\rho}}
\sqrt{\Omega\left(\tilde{\tilde \rho}, l+\displaystyle \frac12\right)} d \tilde{\tilde \rho} \right)}
{ \sqrt[4]{\Omega \left(\rho, l+ \displaystyle \frac12\right) \Omega \left(\tilde \rho, l
+\displaystyle \frac12\right)}}.
\end{eqnarray}
The sum over $l$ can be evaluated by using the Plana sum method (see, for example, paper\cite{Popov})
\begin{eqnarray} \label{plana}
&&\phi(\rho, \theta, \varphi; \tilde \rho, \theta, \varphi)=
\frac{q}{r(\rho) r(\tilde \rho)}
\lim_{\epsilon \rightarrow 0}
\left\{
\int \limits^{\infty}_{\epsilon}
\frac{\exp \left({-\int^{\rho+\delta \rho}_{\rho}}
\sqrt{\Omega(\tilde{\tilde \rho}, x)} d \tilde{\tilde \rho} \right)}
{ \sqrt[4]{\Omega(\rho, x) \Omega(\tilde \rho, x)}} \, x d x
\right. \nonumber \\ && \left.
+\int \limits^{\epsilon}_{\epsilon-i \infty}
\frac{ \exp \left({-\int^{\rho+\delta \rho}_{\rho}}
\sqrt{\Omega(\tilde{\tilde \rho}, z)} d \tilde{\tilde \rho} \right)}
{ \sqrt[4]{\Omega(\rho, z) \Omega(\tilde \rho, z)}
\left( 1+e^{i 2 \pi z} \right)} \, z d z
-\int \limits^{\epsilon+i \infty}_{\epsilon}
\frac{ \exp \left({-\int^{\rho+\delta \rho}_{\rho}}
\sqrt{\Omega(\tilde{\tilde \rho}, z)} d \tilde{\tilde \rho} \right)}
{ \sqrt[4]{\Omega(\rho, z) \Omega(\tilde \rho, z)}
\left( 1+e^{-i 2 \pi z} \right)} \, z d z
\right\}.
\end{eqnarray}
The first integral in this expression can be rewritten as follows
\begin{eqnarray}
&&\frac{q}{r(\rho) r(\tilde \rho)}\int \limits^{\infty}_{0}
\frac{\exp \left({-\int^{\rho+\delta \rho}_{\rho}}
\sqrt{\Omega(\tilde{\tilde \rho}, x)} \ d \tilde{\tilde \rho} \right)}
{ \sqrt[4]{\Omega(\rho, x) \Omega(\tilde \rho, x)}} \, x d x
\nonumber
\end{eqnarray}
\begin{eqnarray}
&&=\frac{q}{\sqrt{r(\rho) r(\tilde \rho)}}\int \limits^{\infty}_{0}
\frac{x \exp\left(-\int^{\rho+\delta \rho}_{\rho}
\sqrt{x^2+\mu(\tilde{\tilde \rho})^2} \ d \tilde{\tilde \rho}/r(\tilde{\tilde \rho})\right)} {\sqrt[4]{x^2+\mu(\rho)^2}\sqrt[4]{x^2+\mu(\tilde \rho)^2}} \, d x
\nonumber
\end{eqnarray}
\begin{eqnarray}
&&=\frac{q}{r(\rho)} \left[1+O\left(\varepsilon_{\mbox{\tiny \sl WKB}}
\frac{\delta \rho}{r} \right) \right] \int \limits^{\infty}_{0}
\frac{x \, d x}
{\sqrt{x^2+\mu(\rho)^2}}
\left[\frac{}{}1+O\left(\varepsilon_{\mbox{\tiny \sl WKB}} m r^2 \delta \rho \right) \right]
\nonumber \\ &&
=\exp\left[-\frac{\displaystyle \sqrt{x^2+\mu(\rho)^2} }{\displaystyle r(\rho)} \delta \rho
+O\left(\varepsilon_{\mbox{\tiny \sl WKB}} \frac{\delta \rho^2}{r^2} \right)
+O\left(\varepsilon_{\mbox{\tiny \sl WKB}} m^2 \delta \rho^2 \right)\right]
\nonumber
\end{eqnarray}
\begin{eqnarray}
&& =\frac{\displaystyle q \exp \left[-\frac{\displaystyle \delta \rho}{\displaystyle r(\rho)} \mu(\rho) \right]}
{\displaystyle \delta \rho}
\left[1+O\left(\varepsilon_{\mbox{\tiny \sl WKB}} \frac{\delta \rho^2}{r^2} \right)
+O\left(\varepsilon_{\mbox{\tiny \sl WKB}} m^2 \delta \rho^2 \right) \frac{}{} \right]
\nonumber
\end{eqnarray}
\begin{eqnarray}
&&=q \left[ \frac{1}{\delta \rho}-\frac{\mu (\rho)}{r(\rho)} \right]
\left[1+O\left(\mu^2 \frac{\delta \rho^2}{r^2} \right)
+O\left(\varepsilon_{\mbox{\tiny \sl WKB}} \frac{\delta \rho^2}{r^2} \right)
+O\left(\varepsilon_{\mbox{\tiny \sl WKB}} m^2 \delta \rho^2 \right) \frac{}{} \right].
\end{eqnarray}
The next two integrals in (\ref{plana}) do not diverge at $\delta \rho \rightarrow 0$
\begin{eqnarray}
 && \lim_{\epsilon \rightarrow 0}
\left\{
\int \limits^{\epsilon}_{\epsilon-i \infty}
\frac{ \exp \left({-\int^{\rho+\delta \rho}_{\rho}}
\sqrt{\Omega(\tilde{\tilde \rho}, z)} d \tilde{\tilde \rho} \right)}
{ \sqrt[4]{\Omega(\rho, z) \Omega(\tilde \rho, z)}
\left( 1+e^{i 2 \pi z} \right)} \, z d z
-\int \limits^{\epsilon+i \infty}_{\epsilon}
\frac{ \exp \left({-\int^{\rho+\delta \rho}_{\rho}}
\sqrt{\Omega(\tilde{\tilde \rho}, z)} d \tilde{\tilde \rho} \right)}
{ \sqrt[4]{\Omega(\rho, z) \Omega(\tilde \rho, z)}
\left( 1+e^{-i 2 \pi z} \right)} \, z d z
\right\}
\nonumber
\end{eqnarray}
\begin{eqnarray}
&&=r(\rho)\lim_{\epsilon \rightarrow 0}
\left\{
\int \limits^{i\epsilon+\infty}_{i\epsilon}
\frac{x d x}{ \sqrt{\mu^2-x^2}\left( 1+e^{ 2 \pi x} \right)}
+ \int \limits^{-i\epsilon+ \infty}_{-i\epsilon}
\frac{x d x}{ \sqrt{\mu^2-x^2}\left( 1+e^{ 2 \pi x} \right)}
+O\left( \delta \rho \right)
\right\}
\nonumber \\ &&
=2 r(\rho) \int^\mu_0 \frac{x d x}{\sqrt{\mu^2-x^2} \left( 1+e^{2 \pi x} \right)}
+O\left( \frac{\delta \rho}{r} \right).
\end{eqnarray}
Thus the zeroth-order WKB approximation of $\phi$ is
\begin{eqnarray}
&&\phi(\rho, \theta, \varphi; \tilde \rho, \theta, \varphi)=
 \frac{q}{\delta \rho}
+\frac{q}{r(\rho)} \left( -\mu(\rho) \frac{}{}
+2 \int^\mu_0 \frac{x d x}{\sqrt{\mu^2-x^2} \left( 1+e^{2 \pi x} \right)}
\right).
\end{eqnarray}

The procedure of the self-force evaluation requires the renormalization
of a scalar potential $\phi(x; \tilde x)$ which is diverged in the
limit $x \rightarrow \tilde x$ (see, for example, papers\cite{1Roth:2004,2Roth:2004}). This renormalization is achieved by subtracting from $\phi(x; \tilde x)$ the DeWitt–Schwinger counterterm $\phi_{\mbox{\tiny \sl DS}}(x; \tilde x)$ and then letting $x \rightarrow \tilde x$\cite{Popov:2011, FroZel:2012}:
\begin{equation} \label{323}
\phi_{\mbox{\tiny \sl ren}}(x)=\lim_{ \tilde x \rightarrow x}
\left[ \phi(x; \tilde x)- \phi_{\mbox{\tiny \sl DS}}(x; \tilde x)\right],
\end{equation}
where
       \begin{eqnarray} \label{phiDS}
        &&\phi_{\mbox{\tiny DS}}(x^i; \tilde x^i)=
        q \left(\frac{1}{\sqrt{2 \sigma}}
        +\frac{\partial g_{t t}(\tilde x)}{ \partial {\tilde x}^i}
        \frac{\sigma^{{i}}}{4g_{t t}(\tilde x)\sqrt{2 \sigma}}
        -m \frac{}{} \right).
        \end{eqnarray}
In this expression
\begin{equation}
\sigma= \frac{g_{i j }(\tilde x)}{2}  {\sigma^i} {\sigma^j}
\end{equation}
is one-half the square of the distance between the points $x$ and $\tilde x$ along the shortest geodesic connecting them and (see, for example, papers\cite{Synge:1960,Popov:2007})
\begin{eqnarray}
        &&{\sigma^i}=-\left(x^i-\tilde x^i\right)
        -\frac12 \Gamma^{i}_{{j}{k}}\left(x^j-{\tilde x^j}\right)\left(x^k-{\tilde x^k}\right)
        \nonumber \\ &&
        -\frac16 \left( \Gamma^{i}_{{j}{m}} \Gamma^{m}_{{k}{l}}
        +\frac{\partial \Gamma^{i}_{{j}{k}}}{\partial {\tilde x^l}}\right)
        \left(x^j-{\tilde x^j}\right)\left(x^k-{\tilde x^k}\right)\left(x^l-{\tilde x^l}\right)
        +O\left(\left(x-{\tilde x}\right)^4\right),\,
        \end{eqnarray}
where Christoffel symbols $\Gamma^{i}_{{j}{k}}$ are calculated at point $\tilde x$.

The DeWitt–Schwinger counterterm $\phi_{\mbox{\tiny \sl DS}}(x; \tilde x)$ in the limit $\theta= \tilde \theta, \varphi=\tilde \varphi$ can be easily calculated using the metric (\ref{metric})
\begin{eqnarray}
&&2 \sigma = \delta \rho^2+O\left(\delta \rho^4\right),
\nonumber \\
&& \phi_{\mbox{\tiny \sl DS}}(\rho, \theta, \varphi; \tilde \rho, \theta, \varphi)=
        =q\left(\frac{1}{\delta \rho}-m+O\left( \frac{1}{L} \right)
        +O\left( \frac{\delta \rho}{L^2} \right) \right).
\end{eqnarray}

Thus $\phi_{\mbox{\tiny \sl ren}}(x)$ is
\begin{eqnarray} \label{phiren}
&&\phi_{\mbox{\tiny \sl ren}}(x)=\lim_{ \delta \rho \rightarrow 0}
\left[ \phi(\rho, \theta, \varphi; \tilde \rho, \theta, \varphi)
- \phi_{\mbox{\tiny \sl DS}}(\rho, \theta, \varphi; \tilde \rho, \theta, \varphi)\right]
\nonumber \\
&=&\frac{q}{r(\rho)} \left(m r(\rho) -\mu(\rho)
+2 \int\limits^{\mu}_0 \frac{x d x}
{\left( 1+e^{2 \pi x} \right)\sqrt{\mu(\rho)^2-x^2} }
\right)
\left(\frac{}{}1+O(\varepsilon_{\mbox{\tiny \sl WKB}})  \right),
\end{eqnarray}
and the single nonzero component of the self-force is
\begin{eqnarray} \label{frho}
&&f_\rho(x)=-\frac{q}{2}\frac{\partial \phi_{\mbox{\tiny \sl ren}}}{\partial \rho}=
\left[
-\frac{q^2}{2r^2}\frac{d r}{d \rho} \left( \mu \frac{}{}
-2 \int\limits^{\mu}_0 \frac{x d x}
{\left( 1+e^{2 \pi x} \right)\sqrt{\mu^2-x^2} }
\right)
\right. \nonumber \\ && \left.
+2 \pi q^2 m^2 \frac{d r}{d \rho}\int\limits^{\mu}_0 \frac{e^{2 \pi x} d x}
{\left( 1+e^{2 \pi x} \right)^2 \sqrt{\mu^2-x^2} } \right]
\left(\frac{}{}1+O(\varepsilon_{\mbox{\tiny \sl WKB}})  \right) \,.
\end{eqnarray}
The functions
\begin{eqnarray}
F(\mu)= \mu \frac{}{}-2 \int\limits^{\mu}_0 \frac{x d x}{\left( 1+e^{2 \pi x} \right)\sqrt{\mu^2-x^2} }
\end{eqnarray}
and
\begin{eqnarray}
G(\mu)= \int\limits^{\mu}_0 \frac{e^{2 \pi x} d x}
{\left( 1+e^{2 \pi x} \right)^2 \sqrt{\mu^2-x^2} }
\end{eqnarray}
can be evaluate numerically.
Let us note that if one uses $ r $ as the new radial coordinate
\begin{equation}
ds^2=-f(r)dt^2+\left(\frac{d \rho}{d r}\right)^2 d r^2+r^2 \left(d\theta^2+\sin^2\theta\, d\varphi^2\right),
\end{equation}
the expression (\ref{frho}) may be rewritten as follows
\begin{eqnarray} \label{fr}
&&f_r= f_\rho \frac{d \rho}{d r}=
\left[-\frac{q^2}{2r^2} F(\mu)
+2 \pi q^2 m^2 G(\mu) \right]
\left(\frac{}{}1+O(\varepsilon_{\mbox{\tiny \sl WKB}})  \right).
\end{eqnarray}

\section{Conclusions}

The considered approach gives the possibility to compute the approximate expression for the self-potential (\ref{phiren}) and the self-force \ref{frho})
on a scalar charge at rest in the spacetime of long throat
(\ref{metric},\ref{lwkb}-\ref{Lm}). Let us note that the used WKB approximation is valid for all the modes (including $l=0$ mode).
This implies also that the approximate solution (\ref{Wsol}) of the equation (\ref{Weq}) does not depend on the conditions at infinity and in considered situation the effect of self-action is a local one even in the limit of massless field.

We also note that the asymptotic behavior of the function $F(\mu)$ for $\mu \gg 1$ is
\begin{figure}[ht]
\vbox{ \hfil \scalebox{0.4} {\includegraphics{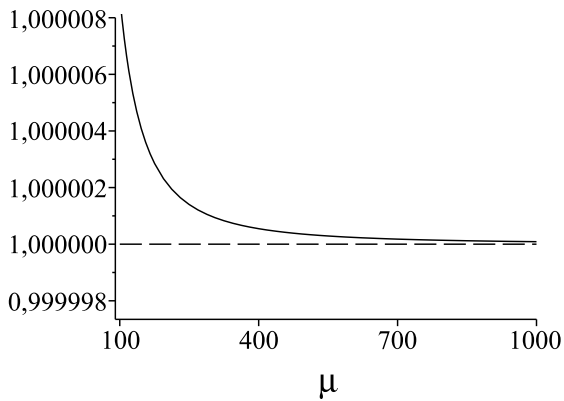}} \hfil }
\caption{The curve represents the function $ \left[\mu-F(\mu)\right] 24 \mu $ for $\mu \gg 1$. }
\end{figure}

This means that the limit of $\phi_{\mbox{\tiny \sl ren}}$ at $ mr \rightarrow \infty $ is equal to (see paper\cite{Popov:2011})
\begin{eqnarray}
\phi_{\mbox{\tiny \sl ren}}(x)&\approx&
         \frac{q}{2 m}\left[ -\frac{{{g_{t t}}_{,{i}}}^{;{i}}}{12 {g_{t t}}}
         +\frac{5{g_{t t}}_{,{i}}{g_{t t }}^{,{i}}}{48 {g_{t t}}^2}
         -\left(\xi -\frac16 \right) R \right]
          -\frac{q}{m r^2}\left(\xi -\frac16 \right)+O\left( \frac{q}{m L^2} \right).
\end{eqnarray}

{\bf Acknowledgments}\\
\noindent
This work was supported in part by grant 13-02-00757
from the Russian Foundation for Basic Research.

\end{document}